%% file: sample-sigconf.tex
\documentclass[sigconf,nonacm]{acmart}

\AtBeginDocument{%
  \providecommand\BibTeX{{%
    \normalfont B\kern-0.5em{\scshape i\kern-0.25em b}\kern-0.8em\TeX}}}

\usepackage[inline]{enumitem}
\usepackage{adjustbox}
\usepackage{graphicx}
\usepackage{algorithm,algpseudocode}

\copyrightyear{2022}
\acmYear{2022}
\setcopyright{acmlicensed}\acmConference[GECCO '22]{Genetic and Evolutionary Computation Conference}{July 9--13, 2022}{Boston, MA, USA}
\acmBooktitle{Genetic and Evolutionary Computation Conference (GECCO '22), July 9--13, 2022, Boston, MA, USA}
\acmPrice{15.00}
\acmDOI{10.1145/3512290.3528826}
\acmISBN{978-1-4503-9237-2/22/07}

\begin{document}

\title{Plasticity and Evolvability Under Environmental Variability: the Joint role of Fitness-based Selection and Niche-limited Competition}

\author{Eleni Nisioti}
\email{eleni.nisioti@inria.fr}
\affiliation{%
  \institution{Flowers Team, Inria and Ensta ParisTech}
  \city{Bordeaux}
  \country{France}
}

\author{Clément Moulin-Frier}
\email{ clement.moulin-frier@inria.fr}
\affiliation{%
  \institution{Flowers Team, Inria and Ensta ParisTech}
  \city{Bordeaux}
  \country{France}
}
\renewcommand{\shortauthors}{Nisioti and Moulin-Frier}

\begin{abstract}

The diversity and quality of natural systems have been a puzzle and inspiration for communities studying artificial life. It is now widely admitted that the adaptation mechanisms enabling these properties are largely influenced by the environments they inhabit. Organisms facing environmental variability have two alternative adaptation mechanisms operating at different timescales: \textit{plasticity}, the ability of a phenotype to survive in diverse environments and \textit{evolvability}, the ability to adapt through mutations. Although vital under environmental variability, both mechanisms are associated with fitness costs hypothesized to render them unnecessary in stable environments. In this work, we study the interplay between environmental dynamics and adaptation in a minimal model of the evolution of plasticity and evolvability. We experiment with different types of environments characterized by the presence of niches and a climate function that determines the fitness landscape. We empirically show that environmental dynamics affect plasticity and evolvability differently and that the presence of diverse ecological niches favors adaptability even in stable environments. We perform ablation studies of the selection mechanisms to separate the role of fitness-based selection and niche-limited competition. Results obtained from our minimal model allow us to propose promising research directions in the study of open-endedness in biological and artificial systems\footnote{We provide code for reproducing results at \url{https://github.com/eleninisioti/ClimateAndLearning} }.

\end{abstract}



\maketitle

\section{Introduction}

\input{sections/intro.tex}

\section{Related works}

\input{sections/related.tex}

\section{Modeling and methodology}

\input{sections/modeling}

\section{Results}

\input{sections/results.tex}

\section{Discussion}
\input{sections/discussion.tex}

\paragraph{Acknowledgements}

This research was partially funded by the Inria Exploratory action ORIGINS (\url{https://www.inria.fr/en/origins}) as well as the French National Research Agency (\url{https://anr.fr/}, project ECOCURL, Grant ANR-20-CE23-0006). This work also benefited from access to the HPC resources of IDRIS under the allocation 2020-[A0091011996] made by GENCI. 

\bibliographystyle{ACM-Reference-Format}
\bibliography{sample-sigconf}

\end{document}

%% file: sections/intro.tex
A key feature of biological evolution is its open-ended nature \cite{darwin,brown_evolutionary_2011}. Understanding how such an apparently simple optimization procedure, based on variation and selection, generates a species with such varying morphological, behavioral and cultural repertoires has been a puzzle for several research communities -- including evolutionary biology, artificial life (AL) and artificial intelligence (AI). A driver of recent progress in our understanding of the emergence of open-endedness in artificial systems is that environments play a key role, complementary to that of the cognitive mechanisms that the AI community has been focusing on for decades \cite{DBLP:journals/corr/abs-2107-12808}. Firm steps towards environments of increased complexity were taken under a family of techniques termed as autocurricula \cite{portelas_automatic_2020}, where aspects of an artificial system such as its multi-agent \cite{baker_emergent_2020,leibo_autocurricula_2019} and curriculum dynamics \cite{wang_paired_2019} are leveraged to automate the emergence of complexity. Generating effective environments is today believed to be one of the pillars for progress in AI \cite{clune_ai-gas_2020}. 

In natural systems, organisms facing environmental variability have two alternative adaptability mechanisms: 
\begin{enumerate*}[label=(\roman*)]
\item \textit{plasti\-city}, the ability of a single genotype to produce multiple phenotypes depending on environmental conditions, enables the survival of the individual under environmental variability within its lifetime \cite{auld_re-evaluating_2010,cuypersEvolutionEvolvabilityPhenotypic2017}. Regulatory homeostasis \cite{cuypersEvolutionEvolvabilityPhenotypic2017}, learning (including its socio-cultural aspect)  \cite{ghalambor_behavior_nodate} and intrinsic motivation \cite{oudeyer_intrinsic_2007} are examples of mechanisms that enable plasticity in natural systems
\item \textit{evolvability}, the ability of modifying the properties of the phenotype through the mutation and selection of genomes. 
\end{enumerate*} Plasticity is intra-ge\-neration; although mechanisms supporting it can be inherited either culturally (e.g. through social norms and language) or genetically (e.g. through morphology and cognitive abilities), behaviors that contribute to plasticity are acquired within one's lifetime and are not transferred by genetic inheritance. Evolvability on the other hand is inter-ge\-neration; it can only act at the moment of reproduction. Both mechanisms are believed to come at a cost. Plasticity is associated with maintenance and production costs, which can be modeled using tolerance curves \cite{auld_re-evaluating_2010,johnston_selective_1982}. Evolvability, on the other hand, increases the probability of deleterious mutations \cite{lynchGeneticDriftSelection2016,doi:10.1126/science.1056421}. Both mechanisms are, thus, evolvable and emerge out of an evolutionary process that optimizes adaptability based an a trade-off of their associated costs and benefits \cite{ghalambor_behavior_nodate,johnston_selective_1982,earl_evolvability_2004}. 

A better understanding of the relationship between adaptability and environment can, thus, be approached by studying the selection pressures that environmental dynamics impose on plasticity and evolvability. Theoretical, data-driven and computational studies have offered many related hypotheses. A consensus seems to be that adaptability disappears in stable environments, as it only incurs costs and no benefits \cite{grove_evolution_2014,cuypersEvolutionEvolvabilityPhenotypic2017}. In particular, the drift-barrier hypothesis attributes the empirical observation that evolvability does not complete disappear but rather converges to a low value to genetic drift caused by the finiteness of population sizes \cite{lynch_lower_2011}. Plasticity on the other hand is hypothesized to emerge when within-generation predictability is not too low to prohibit adaptation and between-generation predictability is not too high for innate mechanisms to suffice \cite{johnston_selective_1982,lange_learning_2021}. Environmental variability in the form of extinction events or periodic variation has been associated with temporary increases in evolvability \cite{lehman_extinction_2015} and plasticity \cite{grove_evolution_2014}.  

Despite this plethora of works, our current understanding of the interplay between environment and adaptability is not complete, arguably due to differences in models and assumptions employed by different studies \cite{lynchGeneticDriftSelection2016,chevinAdaptationPlasticityExtinction2010}. Are the particularities of systems employing adaptability mechanisms so pronounced that one cannot hope for an abstract understanding of adaptability? Our objective with this work is to hint that this is not the case: as we show empirically, even a simple evolutionary model can provide rich insights by taking into account the complex interactions between the heterogeneous mechanisms participating in evolutionary processes: environmental variability, genome properties, fitness-based selection and niche-limited competition. By positioning our observations alongside studies in human ecology and AI, we argue that our understanding of both natural and artificial systems can benefit from such studies.    



In this work we attempt to provide a minimal model of the evolution of plasticity and evolvability and employ it to study its interplay with environmental variability. We study a wide range of  conditions in an environmental model consisting of a climate function and multiple niches. We also analyze the joint effect of fitness-based selection and niche-limited competition and explicitly measure how the trade-off between plasticity and evolvability varies with the selection mechanism. We propose specific measures of phenotypic and genotypic diversity inspired from natural populations. Through an extensive empirical study we show, among others, that:
\begin{enumerate}
    \item Plasticity does not disappear in stable environments under the condition that there is a sufficient number of niches and niche-limited competition.
    \item Both fitness-based selection and niche-limited competition are necessary for survival in complex environments.
    \item Different selection mechanisms give rise to different solutions to the plasticity-evolvability trade-off.
\end{enumerate}

%% file: sections/related.tex
In a recent computational study populations of virtual cells employing a celullar homeostasis mechanism were shown to adapt through evolvability when the frequency of environmental variation was low, while plasticity dominated when variation was frequent \cite{cuypersEvolutionEvolvabilityPhenotypic2017}. While an important step towards distinguishing these two mechanisms, this study did not take into account the presence of niches, studied a specific type of environmental variability and employed a domain-specific genotype to phenotype mapping.

In a minimal model of the evolution of plasticity under environmental variability that follows either a pulse or a sinusoid form  plasticity emerged under environmental variability and completely disappeared during stable periods \cite{grove_evolution_2014}. The mutation rate was kept constant, selection was fitness-based, with individuals competing across niches and a geographical model of niches corresponding to different latitudes was proposed that is also employed in our work. 

The environmental conditions favoring evolvability have been widely studied by the AL community, where accelerating evolutionary optimization can prove vital for practical applications. Major extinctions events, where niches disappear randomly, were linked to increased evolvability \cite{chevinAdaptationPlasticityExtinction2010}. This type of extinction event is unnatural from a biological viewpoint; instead in our work niches disappear indirectly when their capacity, dictated by a climate function, becomes zero and climate varies smoothly with latitude. Reproduction within  limited-capacity niches was shown to also favor evolvability, as evolvable individuals are capable of dispersing, and thus, face a higher effective capacity \cite{lehman_evolvability_2013}. Reproduction within each niche did not take fitness into account, prohibiting the generalization of the conclusion to fitness-based selection mechanisms. Direct search for novelty in behavioral space is also known to favor evolvability \cite{doncieuxNoveltySearchMakes2020}. Instead of explicitly optimizing for novelty, our analysis studies how it emerges as a byproduct of adaptability. In a recent proposal to view evolution as meta-learning, environmental variability and large population sizes were identified as necessary conditions for evolvability to evolve \cite{frans_population-based_2021}, but this study did not take into account the effect of niche-limited competition.

Studies of the effect of environmental dynamics on human evolution, a remarkably plastic species whose behavioral repertoire has inspired our study of artificial systems, point to the fact that early homos were exposed to a mixture of stable, variable and noisy climatic periods \cite{maslin_synthesis_2015,trauth_human_2010}. The rich, multi-scale environmental dynamics arising in such settings are hypothesized to drive adaptability, manifested for example through dispersal patterns, increased mutation rates \cite{harrisEvidenceRecentPopulationspecific2015} and tool use \cite{grove_speciation_2011}. The types of environmental variability examined in our work are largely inspired by the Pulsed Climate Variability framework~\cite{maslin_synthesis_2015}, a notable attempt at unifying existing viewpoints of the interplay between environmental and population dynamics, such as the Red Queen  \cite{doi:https://doi.org/10.1038/npg.els.0001667} and the variability selection hypotheses \cite{potts_hominin_2013}






%% file: sections/modeling.tex
In this section we present our proposed model and methodology for studying the relationship between plasticity and evolvability in variable environments. We separately discuss our modeling of the environment and genome, the evolutationary algorithm and the metrics used to evaluate the population.


\begin{figure*}
\begin{minipage}{0.4\textwidth}
    \includegraphics[width=0.8\textwidth]{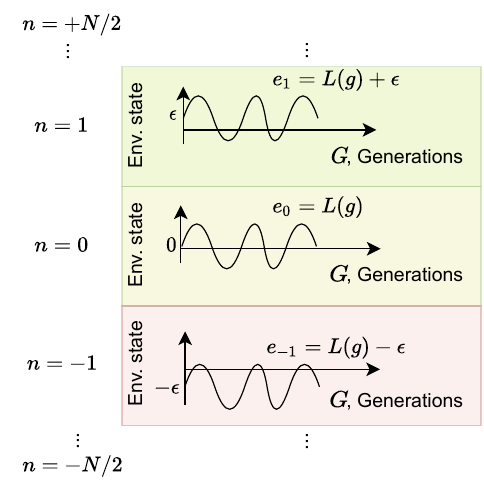}
\end{minipage} \hfill
\begin{minipage}{0.55\textwidth}
    \includegraphics[scale=0.9]{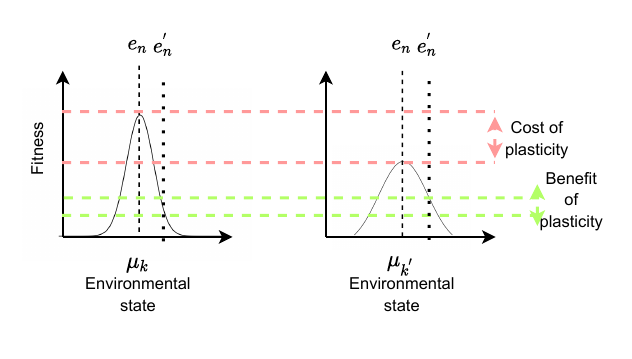}
\end{minipage} 
\caption{(Left) The latitudinal model we employ to describe how the environmental state varies across niches: a single climate function "(sinusoidal curve $e_n$)" $L$ evolves identically for each niche and has a vertical offset equal to $\epsilon \cdot n$ for each $n$. Thus niches with higher index $n$ have higher states, and therefore, higher capacity. (Right) Modeling plasticity as a normal distribution $\mathcal{N}(\mu_k,\sigma_k)$. A non-plastic individual ($k$) has small $\sigma_k$ and a high peak at their preferred niche, while a plastic individual ($k^{'}$) has large $\sigma_k$ and a lower peak at their preferred niche. Fitness in a given niche $n$ is computed as the probability density function of the distribution at the environmental state $e_n$. This figure also illustrates the cost and benefit of plasticity, assuming that $\mu_k=\mu_k^{'}$. If $e_n=\mu_k$ (the actual environmental state is identical to the preferred niche of both individuals) the plastic individual has lower fitness (cost of plasticity). If $e_n >>\mu_k$ (the actual environmental state differs significantly from the preferred one) the plastic individual has higher fitness (benefit of plasticity).}
 \label{fig:tol_curve}
\end{figure*}

A population of $\mathcal{K}=\{1,\ldots, K\}$ individuals evolves in an environment consisting of $\mathcal{N}=\{1,\ldots, N\}$ niches for $\mathcal{G}=\{1,\ldots, G\}$ generations under selection mechanism $S$ and genome model $O$.

\subsection{Modeling the environment}\label{sec:env_model}
An environment is characterized by: 
\begin{enumerate*}[label=(\roman*)]
    \item a number of niches $N$ arranged in a simple latitudinal model: we consider a reference niche at $n=0$, $N/2$ ``northern" niches indicated with positive indexes $n \in (0,N/2]$  and $N/2$ ``southern" niches with negative indexes $n \in [-N/2,0)$ 
    \item a climate function $L: \mathcal{G} \rightarrow \mathbb{R} $, which describes the value of the environmental state of the reference niche $n=0$  at a given generation $g$ , i.e. $e_0^g=L(g)$. 
\end{enumerate*} The environmental state of any niche $n$ can then be determined as:
\begin{align}\label{eq:niche}
    e_n^g=e_0^g + n \cdot \epsilon,
\end{align} with $\epsilon$ a constant capturing the difference in terms of environmental state between adjacent niches. The schematic on the left of Figure \ref{fig:tol_curve} illustrates how the environmental state of each niche is computed based on our model.

The environmental state determines the fitness of the individual based on its genome, as described in Section \ref{sec:genome_model} and  the capacity of a niche $c_n^g$ as: $c_n^g=e_n^g C_N$, where $C_N$ is termed the climate-independent capacity. In order to ensure that the maximum population size is independent of the number of niches we define $C_N=C_{\text{ref}}/N$, where $C_{\text{ref}}$, the reference capacity, is equal to the desirable maximum population size. Thus, higher environmental states can support larger populations and are termed ``high-quality": for a given climate function, ``northern" niches have higher quality, while environments where the climate function takes higher values have higher quality in general.  An assumption of this model is that there is spatial smoothness, i.e, nearby niches are more similar.

\subsection{Modeling the genome}\label{sec:genome_model}
We adopt tolerance curve to design a genome that can track both directional selection and selection for plasticity \cite{doi:10.1086/284635,grove_evolution_2014}. A tolerance curve is a normal distribution with mean $\mu_k^g$, indicating the environmental state of highest fitness for an individual, which we refer to as the preferred state, and a standard deviation $\sigma_k^g$ that captures how quickly the fitness of the genome drops as the environmental state varies from its preferred state. Genomes with large $\sigma_k^g$ are indicative of plastic individuals, as these can tolerate a larger variety of states. On the right of Figure \ref{fig:tol_curve} we can see the tolerance curves of a plastic and a non-plastic individual. As the sum of the area under the curve is always equal to one, this model captures the cost and benefit of plasticity. The genome $o_k^g$ also includes the mutation rate $r_k^g$, thus $o_k^g=[\mu_k^g, \sigma_k^g, r_k^g]$. Upon reproduction the genome mutates as:

\begin{align} 
\mu_k^{g+1} = \mu_k^{g} + \mathcal{N}(0, r_k^{g}) \nonumber \\
\sigma_k^{g+1} =\sigma_k^{g} + \mathcal{N}(0, r_k^{g}) \nonumber \\
r_k^{g+1} = r_k^{g} +\mathcal{N}(0, r_k^{g}) \label{eq:mutation}
\end{align}
where $\mathcal{N}(x,y)$ denotes a normal distribution with mean $x$ and variance $y$. This genome model, which we refer to as $R_{\text{evolve}}$, captures the co-evolution of plasticity (through $\sigma$) and evolvability (through $r$). In our experiments, we also try a simplification of this model, $R_{\text{no-evolve}}$, where the mutation rate is constant (i.e., $r_k^g=r_0 \forall g, \forall k$).

\subsection{Selection mechanism}\label{sec:select_mechanism}
At the end of a generation individuals are selected for sexual reproduction based on the selection mechanism $S$ and their offspring form the next generation. To compute the fitness of an individual $k$ in generation $g$ we first detect the niches in which it can survive as: 
\begin{align}\label{eq:survive_niche}
{n \in \{1, \cdots, N\} \quad | \quad e_n^g \in [\mu_{k}^g - 2\sigma_{k}^g,\mu_k^g + 2\sigma_k^g]}
\end{align} and
compute its fitness in each one of them as $f_{k,n}^g=pdf(\mu_k^g, \sigma_k^g, e_n^g)$, where $pdf$ denotes the value of the normal probability density function with mean $\mu_k^g$, and  variance $\sigma_k^g$ at location $e_n^g$. 

Our proposed selection mechanism entails two independent assumptions inspired from natural evolution:
\begin{enumerate*}[label=(\roman*)]
\item \textit{niche-limited competition}: when deciding which individuals will reproduce, we study each niche independently;
\item \textit{fitness-based selection}: within a niche, individuals produce offspring until its capacity is filled, with fitter individuals being chosen with higher probability
\end{enumerate*}. Only individuals that can survive in a niche are considered for reproduction within it. We refer to this mechanism as NF-selection.

We also experiment with two ablations of this mechanism:
\begin{enumerate*}
\item under F-selection, competition is population-wide and individuals are selected for reproduction based on their fitness. This model is often termed as survival of the fittest. 
\item under N-selection, individuals reproduce only within their own niche and are chosen with equal probability until its capacity is filled. This model is commonly known as limited-capacity.
\end{enumerate*} 

We provide the pseudocode describing how evolution takes place in Algorithm \ref{alg:evolution}, which explain how the environment and population change with generations based on the climate function $L$, and Algorithm \ref{alg:reproduce}, which provides more details on how reproduction differs depending on the selection mechanism.

\subsection{Evaluation metrics}\label{sec:eval_metrics}
To evaluate the population we compute the following metrics at the end of each generation $g$:
\begin{enumerate}
    \item $\bar{\mu}^g$, the value of the preferred environmental state, averaged over the population. This metric indicates that the population is well-adapted when $\bar{\mu}^g$ tracks the form of the climate function. Insensitivity of $\bar{\mu}^g$ to environmental change comes at a fitness cost due to our genome model. 
    \item $\bar{\sigma}^g$, the value of the standard deviation $\sigma^g$ for preferred environmental states, averaged over the population. We refer to this metrics as the population-average plasticity.
    \item $\bar{r}^g$, the mutation rate $r$ component averaged over the population, which denotes the population-average evolvability.
    \item $X^g=\sum_k X_k^g$, the number of extinctions. We denote the survival of individual $k$ in niche $n$ at generation $g$ as a binary variable:
    \begin{align}\label{eq:survival}
        s_{k,n}^g = (e_{n,g} \in [\mu_{k}^g - 2\sigma_{k}^g,\mu_k^g + 2\sigma_k^g])
    \end{align}
    Thus, an individual goes extinct ($X_k^g=1$) if $\sum_n^N s_{k,n}^g$ is zero and survives ($X_k^g=0$) if $\sum_n^N s_{k,n}^g$ is positive. 
    \item population survival $A^g$, the percentage of generations that a run of our algorithm survived for. Values smaller than 1 are indicative of a mass extinction.
    \item $V^g$, the diversity of the population defined as the standard deviation of the population's genes, formally:
    \begin{align}\label{eq:diversity}
        V = \sigma_{\mu^g} + \sigma_{\sigma^g} +  \sigma_{r^g}
    \end{align} 
    This metric captures the genetic diversity of the population.
    \item $D^g$, the dispersal of the population, computed as the number of niches over which at least one individual survives for a temporal window of at least $w$ generations. Formally, $D^g=\sum_{n=1}^N d_{n,w}^g$,  where $d_{n,w}^g$ denotes the persistence of the population in a given niche for the required time window and is computed as
    \begin{align}
        d_{n,w}^g = \begin{cases} 1 \quad \text{if} \sum_{g^{\prime}=g-w}^g s_{n}^{g^{\prime}}=w \\
        \text{0} \quad \text{otherwise}
        \end{cases}
    \end{align}
    where $s_{n}^g$ is indicates the survival of at least one individual in a given niche and is computed as 
    \begin{align}
        s_{n}^g = \begin{cases}
        1 \quad {if} \sum_{k}^K s_{k,n}^g > 1 \\
        0 \quad \text{otherwise}
        \end{cases}
    \end{align} with $s_{k,n}^g$ defined in Eq. (\ref{eq:survival}). As this metric arises from the interaction of the genome and environment, we can view it as a measure of phenotypic diversity. 
    
\end{enumerate}

Table \ref{tab:notation}  contains a summary of our notation for parameters characterizing the population, environment and evaluation metrics .

%% file: sections/results.tex
We study the behavior of a population following our proposed model of NF-selection with the $R_{\text{evolve}}$ genome in a variety of environments that differ in the form of the climate function $L$, which is either stable, sinusoidal or noisy and the number of niches $N$, sampled in the range $(1,100)$. The reference capacity is $C_{\text{ref}}=1000$ and the distance between niches is $\epsilon=0.01$. We denote values after convergence with an asterisk super-script. We perform ablation studies of the selection mechanism by also experimenting with F-selection and N-selection and the genome model, by comparing to $R_{\text{no-evolve}}$. For each experiment we present the mean of the evaluation metrics and $95\%$ confidence intervals computed over 20 independent runs.

\subsection{Evolving in a stable environment}
We define a stable environment as one where the climate function, and hence, the reference environmental state is constant, i.e. $e_0^g=e_0^{0}, \quad \forall g \in \mathcal{G}$. The environmental state of the different niches is, therefore, equal to $e_n^g=e_0^{0} + n \cdot \epsilon$. We experiment with high-quality environments ($e_0^{0} > 4$) which can support a large population, low-quality environments ($ e_0^{0} \leq 0.5$) where the capacity is low with the majority of niches being unable to support any individuals and medium-quality environments (0.5 < $e_{0}^0 < 4$) where some of the niches become uninhabitable for large enough $N$.

\begin{figure}
    \centering
    \includegraphics{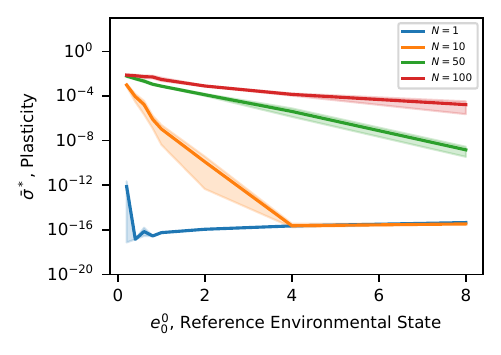}
    \caption{Population-average plasticity after convergence ($\bar{\sigma}^*$) in a constant environment under NF-selection and genome $R_{\text{evolve}}$.}
    \label{fig:stable_sigma}
\end{figure}

\subsubsection{Low-quality environments with multiple niches favor plasticity}\label{sec:stable_plasticity}

Figure \ref{fig:stable_sigma} presents the population-average plasticity after convergence, $\bar{\sigma}^*$, under NF-selection using the $R_{\text{evolve}}$ genome model (populations converged after around 100 generations) under various environmental conditions and number of niches. We observe that when there is a single niche ($N=1$) plasticity converges to a very low value regardless of the state. This is intuitive as the cost of plasticity captured by our genome model renders individuals with the smallest $\sigma_k^*$ the fittest. This agrees with previous studies in constant environments \cite{grove_evolution_2014} (note that F-selection is identical with NF-selection when there is a single niche). However, the picture differs significantly when there are multiple niches and low-quality environments: as an individual can reproduce in any of the niches it can survive in, higher plasticity means higher chances of reproduction, which counteracts the cost of plasticity. As the quality of environments increases the benefit of plasticity disappears: non-plastic individuals dominate the available niches even though some individuals choose to disperse. In contrast to these interesting dynamics of plasticity, we observed that the population-average evolvability remained very low ($\bar{r}^* < 10^{-10}$) in all conditions. This observation is inline with the intuition that mutations disappear in stable environments as they incur fitness costs \cite{lynchGeneticDriftSelection2016,doi:10.1126/science.1056421}.

\subsubsection{Niche-limited competition is necessary for plasticity to persist}\label{sec:stable_plasticity}

\begin{figure}
    \centering
    \includegraphics{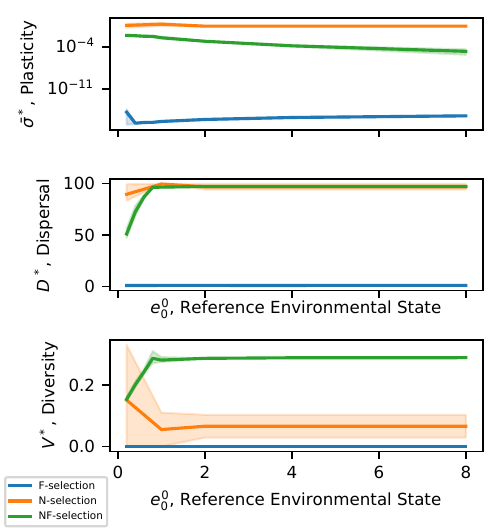}
    \caption{Population-average plasticity ($\bar{\sigma}^*$), dispersal ($D^*$) and diversity ($V^*$) after convergence in a constant environment under different selection mechanisms and environmental conditions for $N=100$ niches: F-selection leads to considerably lower plasticity, dispersal and diversity.}
    \label{fig:stable_selection}
\end{figure}

In Figure \ref{fig:stable_selection} we compare how the different selection mechanisms behave under different environmental conditions when there is a large number of niches ($N=100$). We observe that under F-selection plasticity and dispersal are very low, while N-selection and NF-selection exhibit a similar behavior. This suggests that a key element in maintaining plasticity in a constant environment is combining a large number of niches with niche-limited competition and explains why our conclusions disagree with existing studies in constant environments that only consider F-selection in a single niche \cite{grove_evolution_2014}. Also, the high dispersal under N-selection agrees with studies hinting that niche-limited competition that ignores fitness leads to high phenotypic diversity \cite{lehman_evolvability_2013}. However, our study further clarifies that, when given an option to adapt through plasticity or evolvability, both N-selection and NF-selection opt for plasticity. This phenomenon was not captured by models that considered evolvability as the only adaptability mechanism \cite{lehman_evolvability_2013}.

\subsubsection{Diversity is highest under NF-selection}\label{sec:stable_diversity}

In Figure \ref{fig:stable_selection} we observe that diversity $V^*$, defined in Eq. \ref{eq:diversity}, under NF-selection increases as the environmental state increases, while the opposite happens for N-selection, which also exhibits higher variance. Under F-selection, the population exhibits no diversity. Thus, both fitness-based selection and niche-wide competition are necessary for diversity to be high. A closer inspection of the constituents of diversity reveals that the high diversity of NF-selection is due to the $\sigma_{\mu^g}$ component. Specifically, $ \sigma_{\mu^*}=0.2849,\sigma_{\sigma^*}=0.0015, \sigma_{r^*}=4 \cdot 10^{-7} $ for NF-selection and $ \sigma_{\mu^*}=0.01,\sigma_{\sigma^*}=0.00243, \sigma_{r^*}=4 \cdot 10^{-10}$ for N-selection. Thus, NF-selection and N-selection may both exhibit the same dispersal (see Fig \ref{fig:stable_selection}) but their reasons are different: for NF-selection it's a combination of high plasticity ($\bar{\sigma}$) and high diversity in the preferred niches ($\sigma_{\mu^g}$), while for N-selection it is solely due to the, slightly higher than under NF-selection, plasticity. 

\subsubsection{F-selection leads to more early extinction events}\label{sec:stable_extinct}

\begin{figure}
    \centering
    \includegraphics{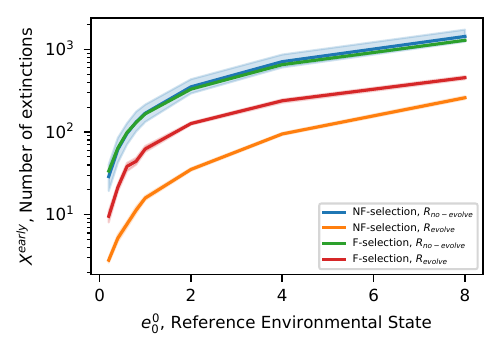}
    \caption{Extinction events ($X^*$) in a constant environment under different selection mechanisms and genome models for $N=100$ niches.}
    \label{fig:stable_extinct}
\end{figure}

In Figure \ref{fig:stable_extinct} we monitor early extinction events, $E^{early}$ defined as the average number of extinctions during the first 100 generations. We observe that populations under F-selection suffer much higher extinction rates compared to populations under NF-selection .  This is the case only for evolvable populations ($R_{\text{evolve}}$): if we don't allow evolvability to disappear but instead keep it constant at $r_k^g=0.001 \forall g, k$, then extinctions persist both for F-selection and NF-selection. The behavior for N-selection is similar to NF-selection suggesting that high extinctions in evolvable populations are due to lack of niche-limited competition: indeed as we saw in Figure \ref{fig:stable_selection}, individuals under F-selection have very low plasticity ($<10^{-10}$) and occupy a single niche. Thus, mutations, even though occurring with low probability can slowly shift an individual away from its niche and lead to its extinction. Note that the distance between niches is too large ($\epsilon=0.01$) for a mutation to transfer a non-plastic individual to another niche. The trend of increasing extinctions with the reference environmental state is consistent for different values of $N$ and is a result of the fact that the population size increases. In general we observed that all extinction events happen at the early stages of evolution ($g<100$), except for settings under NF-selection and high climate values. In these cases we observed that extinctions persist until generation $g=1000$, which suggests that the population required more time to converge, although it still experience less early extinctions that F-selection.

\subsection{Evolving under periodic variability}
We model periodic variability as a sinusoid with period $T_e$ and amplitude $A_e$ that dictates the evolution of the reference environmental state $e_{0,g}$. Based on our environmental model described in Section \ref{sec:env_model} and illustrated in Figure \ref{fig:tol_curve}, niches experience changes simultaneously but the exact environmental state has an offset that depends on the latitude. Compared to the previous analysis of stable environments, behaviors here exhibit more rich dynamics. To better analyze them, we first present results on survival $A$ (defined in Section \ref{sec:eval_metrics}) and then monitor how specific metrics evolve with generations.  

\subsubsection{Plastic and evolvable individuals emerge under NF-selection only when the number of niches is sufficient}\label{sec:periodic_NF}

\begin{figure}
\centering
    \includegraphics{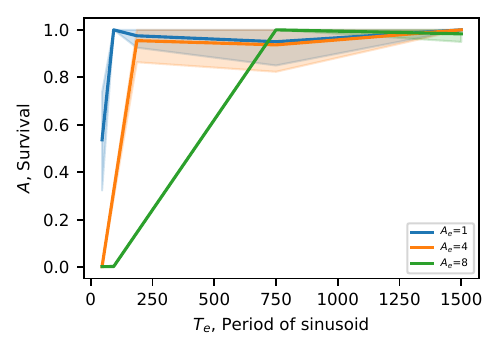}
    \caption{Survival ($A$) as the percentage of generations without a mass extinction under NF-selection with genome $R_{\text{evolve}}$, $N=$100 niches and varying period $T_e$ and amplitude $A_e$.}
    \label{fig:sin_survival_A}
\end{figure}

\begin{figure}
\centering
    \includegraphics{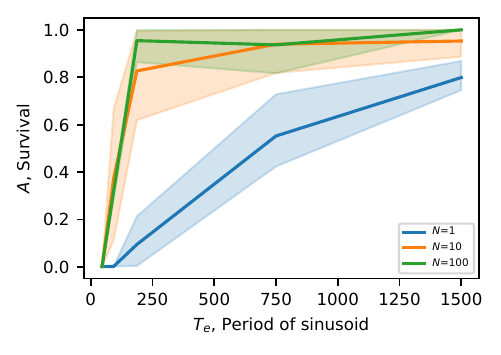}
    \caption{Survival  ($A$) as the percentage of generations without a mass extinction under NF-selection with genome $R_{\text{evolve}}$, $A_e=4$  and varying period $T_e$ and Number of niches.}
    \label{fig:sin_survival_N}
\end{figure}

In Figures \ref{fig:sin_survival_A} and \ref{fig:sin_survival_N} we observe the ability of the population to survive when we vary the amplitude of oscillations ($A_e$) and number of niches ($N$) respectively, for different values of the oscillation period $T_e$. We can draw various conclusions from these results:
\begin{enumerate*}[label=(\roman*)]
\item increasing the number of niches enables the population to survive longer in environments changing more frequently
\item survival is guaranteed for small-amplitude variations ($A_e=0.2$) regardless of their frequency
\item in the case of large-amplitude variations ($A_e=8$)  high frequency does not allow the population to adapt at all
\item In the case of medium-amplitude ($A_e=1$) survival is possible only under low-frequency variation 
\end{enumerate*}

\begin{figure}
\centering
    \includegraphics{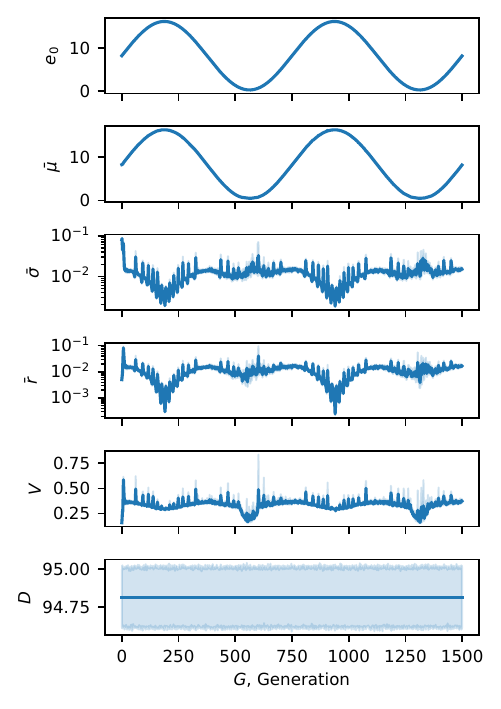}
    \caption{Evolution under NF-selection with genome $R_{\text{evolve}}$, $N=$100 niches, $T_e=750$ and $A_e$=8.}
    \label{fig:sin_evolution_slow}
\end{figure}

In Figure \ref{fig:sin_evolution_slow}, we further analyze the ability of the population to survive by looking at how the evaluation metrics evolve with increasing generations. We observe that the population manages to track the environmental variability by keeping both plasticity and evolvability high, with oscillations in plasticity and evolvability occuring at twice the period of variability $T_e$, as they increase at both transition points. Diversity is slightly lower during the low peaks of $e_0$. We should note that we did not find high-frequency cases where the population reacted solely through plasticity or low-frequency cases where the population adapted solely through evolvability, as indicated by previous studies \cite{cuypersEvolutionEvolvabilityPhenotypic2017}.

\subsubsection{Maladapted  plastic individuals emerge under F-selection and N-selection}
\begin{figure}
\centering
    \includegraphics{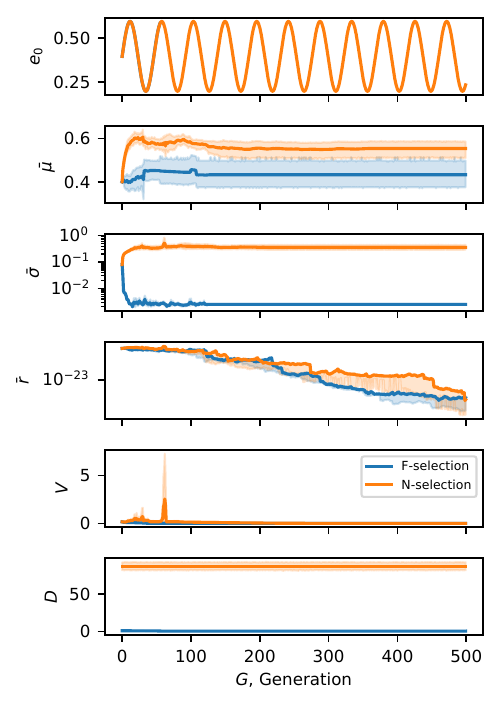}
    \caption{Evolution under F-selection (blue) and under N-selection (orange) with genome $R$, 100 niches, $T_e=46$ and $A_e$=0.2.}
    \label{fig:sin_evolution_quick}
\end{figure}

To understand the effect of the selection mechanism on the course of evolution, we present similar results under F-selection and N-selection in Figure \ref{fig:sin_evolution_quick}. As these mechanisms can avoid mass extinctions only for very small amplitudes and periods, we present results for $A_e=0.2$ and $T_e=46$. Low frequency settings with high amplitude can be paralleld to major extinction events \cite{lehman_extinction_2015}, where N-selection and F-selection fail due to their low evolvability.   Figure \ref{fig:sin_evolution_quick} further shows some similarities between the two mechanisms: in both of them evolvability becomes very small and adaptation essentially stops: we see that $\bar{\mu}$ does not track the sinusoid. Also, diversity is very low. On the other hand, the two mechanisms exhibit significant differences: under F-selection there is no dispersal and plasticity is low. Under D-selection, dispersal is significantly higher ($D=75$) and plasticity is higher. This suggests that under F-selection the population survives because specialists occupy a few niches, while under N-selection the population survives by occupying as many niches as possible.

\subsection{Evolving in a noisy environment}
In a noisy environment the reference state $e_0^g$ takes values sampled from a normal distribution with standard deviation $\sigma_N$ and mean $e_0^0$. We experiment with various environmental conditions with $\sigma_N \in (0.05,0.8)$ and $e_0^0 \in (0.2,4)$. 

\subsubsection{Under niche-limited competition populations can tolerate higher uncertainty}\label{sec:noisy_NF}
\begin{figure}
\centering
    \includegraphics{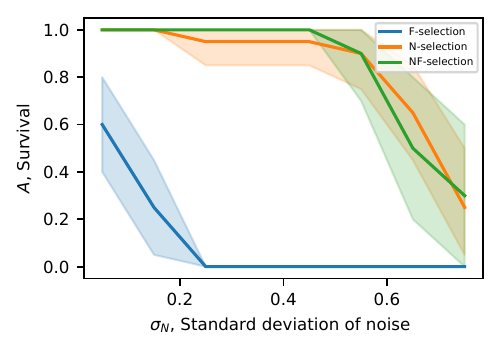}
    \caption{Survival under different selection mechanisms and varying noise levels for $e_0^0=2$ and $N=100$ niches with genome $O_{\text{evolve}}$.}
    \label{fig:noisy_survival}
\end{figure}

In Figure \ref{fig:noisy_survival} we present how the survival ability of the population varies with the level of noise under different selection mechanisms for an environment with $N=100$ niches and $e_0^0=2$. Under the same environmental conditions for $N=1$ we observed only extinctions. This is because a single niche prohibits the survival of plastic individuals due to the cost of plasticity. We also observed that for $N=100$ and $e_0=0.2$ populations under NF-selection and N-selection survived only for the minimum examined noise level ($\sigma_N=0.05$). This is not surprising as in uncertain low-quality environments niches disappear (their capacity becomes 0) with high probability so that dispersal cannot offer an advantage. 

\begin{figure}
\centering
    \includegraphics{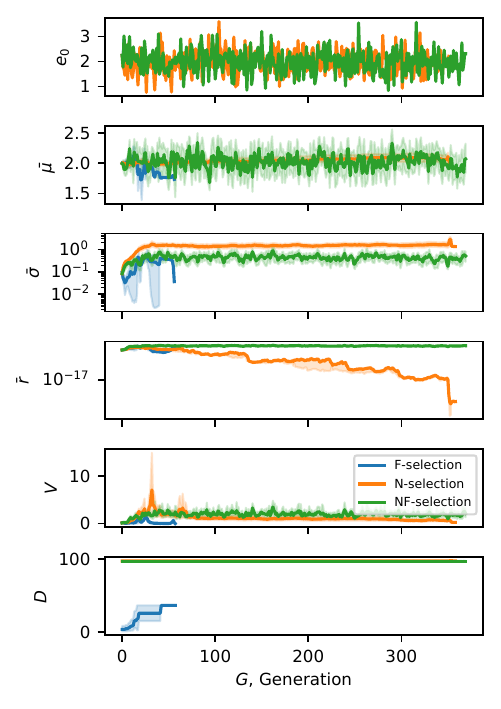}
    \caption{Evolution under F-selection (blue), N-selection (orange) and NF-selection (green)  with genome $R_{\text{evolve}}$, $N=40$ niches, $\sigma=0.2$ and   $e_0^0=2$.}
    \label{fig:noisy_evolution}
\end{figure}

In Figure \ref{fig:noisy_evolution} we monitor different selection mechanisms for $N=100$ niches, $\sigma=0.2$ and   $e_0^0=2$. We observe that:
\begin{enumerate*}[label=(\roman*)]
\item under F-selection a mass extinction occurs early on due to low plasticity
\item under N-selection plasticity is high and evolvability low. The population does not track the variations of $e_0$ and diversity is low
\item under NF-selection the population has both high plasticity and evolvability, as well as higher diversity. 
\end{enumerate*} Thus, in line with other studies \cite{cuypersEvolutionEvolvabilityPhenotypic2017}, we find settings where plasticity and evolvability behave differently (N-selection). We observe that to deal with high uncertainty, a population needs to either keep both plasticity and evolvability high (NF-selection) or reduce evolvability and require higher values of plasticity (N-selection). Thus, fitness-based selection introduces an upper limit to plasticity (due to the plasticity cost) which forces the population to adapt through its evolvability. %

%% file: sections/discussion.tex
We have designed a simple model of the evolution of plasticity and evolvability and studied the complex interactions between environmental and population dynamics. Despite its simplicity, experiments have revealed many insights into the evolution of adaptation mechanisms. Taking into account the effect of niche-limited competition gives rise to qualitatively different solutions to the plasticity-evolvability trade-off, in turn affecting population properties such as diversity and dispersal. We hope that our work sheds light into the plethora of related works and will prove useful in future studies of both artificial and natural systems.

We believe that Quality-Diversity algorithms \cite{pugh_quality_2016} can be a particularly promising application area for such studies. Similarly to our proposal, this community lays emphasis on the benefits of combining niche-limited competition and fitness-based selection; however, as our empirical results indicate, parameters such as the number and quality of niches, as well as the form and presence of environmental variability show great qualitative impact and can potentially act as a curriculum for the emergence of adaptation.

From the perspective of human ecology our observation show that such simple can offer insights into existing hypotheses: 
\begin{enumerate*}[label=(\roman*)]
\item the observation that low-quality environments favor plastic individuals, while high\-quality environments favor non-plastic individuals (see Section \ref{sec:stable_plasticity}) hints to the turnover pulse hypothesis \cite{Vrba1985EnvironmentAE};
\item the observation that adaptability is favored by abrupt transitions (see Section \ref{sec:periodic_NF}) and high variability (see Section \ref{sec:noisy_NF}) hint to the variability selection hypothesis \cite{potts_hominin_2013}
\end{enumerate*} We should note that these hypotheses share a common model of environmental variability characterized by a large diversity of niches \cite{maslin_synthesis_2015}. 

It is important to also note the assumptions made by our study and how future work in AI and AL can help overcome its limitations. First, tolerance curves assume that plasticity comes at a cost, an assumption that is often questioned in natural systems; we believe that studies with artificial systems can reveal whether such costs indeed arise. Second, our model does not capture the mechanism of species co-adaptation \cite{sole_revisiting_2022} and can therefore not offer insights on how the dynamics of arms races are influenced by resource availability, as proposed by the Red Queen hypothesis \cite{doi:https://doi.org/10.1038/npg.els.0001667}. Then, our model assumes that there are no constraints on plasticity; we believe that studies with a more complex genotype to phenotype mapping that employ different adaptation mechanisms to ensure plasticity can reveal mechanism-specific limits that will extent the conclusions reached in this work. From an evaluation perspective, we have limited ourselves to measuring easily quantifiable properties of the genome and behavior space, such as adaptability and diversity, that have been linked to open-endedness \cite{pugh_quality_2016}. As a next step we plan to investigate direct measures of open-endedness \cite{Dolson2019TheMT}. Finally, we believe that progress in open-endedness requires a better understanding of niching in recent simulation environments employed by the deep reinforcement learning community \cite{suarez_neural_2019,pmlr-v97-cobbe19a,nisioti:hal-03446961}. 

Finally, as suggested by a conceptual framework for modeling open-ended skill acquisition in artificial and natural systems \cite{nisioti:hal-03446961}, the effect of environmental variability on adaptation mechanisms is only part of the overall picture. To understand the  complexity of the human ecological niche, we need to take into account multi-agent dynamics and culture, which modulate the processes of selection and niche construction \cite{eppe:hal-03120583,nisioti:hal-03446961}. We believe that exploring these links in social-centric multi-agent reinforcement learning environments \cite{DBLP:journals/corr/abs-2104-13207} will further improve our understanding of adaptability.